# Kaon Interferometry

U. Heinz[1,a], B. Tomášik[2], U.A. Wiedemann[2] and Wu Y.-F.[2,b]

[1] Physics Department, Duke University, Durham, NC 27708-0305, USA
[2] Institut für Theoretische Physik, Universität Regensburg,
D-93040 Regensburg, Germany

**Abstract.** 2-kaon and 2-pion correlation functions for an expanding thermalized source are compared. In the Yano-Koonin-Podgoretskii parametrization of the correlation function, the HBT radius parameters are shown to obey $M_\perp$-scaling in the absence of collective transverse flow. This scaling is broken by transverse flow. An accurate comparison of pion and kaon correlations can thus resolve issue whether the observed $M_\perp$-dependence of the transverse radius parameter is due to transverse collective flow or other transverse gradients. Effects from resonance decays are shortly discussed.

## 1. Introduction

In the last few years a large body of evidence has been accumulated that the hot and dense collision region in ultrarelativistic heavy ion collisions thermalizes and shows collective dynamical behaviour. This evidence is based on a comprehensive analysis of the hadronic single particle spectra. It was shown that all available data on hadron production in heavy ion collisions at the AGS and the SPS can be understood within a simple model which assumes locally thermalized momentum distributions at freeze-out, superimposed by collective hydrodynamical expansion in both the longitudinal and transverse directions [1]-[6]. The collective dynamical behaviour in the transverse direction is reflected by a characteristic dependence of the inverse slope parameters of the $m_\perp$ spectra ("effective temperatures") at small $m_\perp$ on the hadron masses [1, 2]. New data from the larger Au+Au and Pb+Pb systems [7, 8] support this picture and show that the transverse collective dynamics is much more strongly exhibited in larger collision systems than in the smaller ones from the first rounds of experiments. The amount of transverse flow appears also to increase monotonically with collision energy from GSI/SIS to AGS energies, but may show signs of saturation at the even higher SPS energy [9].

It must be noted, however, that the "proof" of thermal and hydrodynamic behaviour from an analysis of single particle spectra strongly relies on model assumptions. A model independent extraction of flow velocities and thermal freeze-out temperatures from the measured momentum spectra is not possible [2, 4]. Explicit dynamical model calculations and additional theoretical consistency arguments [3, 10, 11] are required to achieve this goal, and there have been numerous alternative suggestions to explain the observed features of the hadron spectra without invoking hydrodynamical flow, but using differently motivated parametrizations [12]. The reason for this ambiguity is that single particle momentum spectra, as a matter of principle, contain no direct information on the space-time structure and the space-momentum correlations induced by collective flow [13]: In terms of the phase space density at freeze-out ("emission function") $S(x,p)$ the single-particle spectrum is given by $E\,dN/d^3p = \int d^4x\, S(x,p)$, and one sees that the space-time information contained in $S$ is completely washed out by integration. Thus, on the single-particle level, one must perform comprehensive model analyses to show that a simple hydrodynamical model with only a few thermodynamic and collective parameters can fit all the data, and one must use additional consistency checks in order to show that the values of the extracted fit parameters lead to a sensible and internally consistent theoretical picture. The published literature abounds with examples demonstrating that without such consistency checks the theoretical ambiguity of the single particle spectra is nearly infinite.

This is the point where two-particle Bose-Einstein correlations between the momenta of pairs of identical particles provide crucial new input. They give direct access to the space-time structure of the source and its collective dynamics, and although some remaining model dependence cannot be avoided the set of possible model sources is drastically reduced. The two-particle correlation function $C(q,K)$ is usually well approximated by a Gaussian in the relative momentum $q$, with Gaussian width parameters ("HBT (Hanbury-Brown/Twiss) radii") which depend on the total momentum $K$ of the particle pair. It was recently shown [14]-[17] that these radius parameters measure certain combinations of the second order variances of the space-time structure of the source (the so-called "regions of homogeneity" [18] of the source around its point of maximum emission). In general they mix the spatial and temporal structure of the source in a nontrivial way [15, 16], and the mentioned model dependence enters when trying to unfold these two aspects.

Collective dynamics of the source leads to a dependence of the HBT radii on the pair momentum $K$; this has been known for many years [19, 20, 21], but was recently quantitatively reanalyzed, both analytically [22, 16, 17, 23, 18] and numerically [24, 13, 25]. The velocity gradients associated with collective expansion lead to a dynamical decoupling of different source regions in the correlation function, and the HBT radii measure the size of the resulting space-time regions of homogeneity of the source [21, 18] around the point of maximum emissivity for particles with the measured momentum $K$. The velocity gradients are smeared out by a thermal smearing factor arising from the random motion of the emitters around the fluid velocity [16]. Due to the exponential decrease of the Maxwell-Boltzmann distribution, this smearing factor shrinks with increasing transverse momentum $K_\perp$ of the pair, which is the basic reason for the $K_\perp$-dependence of the HBT radii.

Unfortunately, other transverse and longitudinal gradients in the source (for example spatial and temporal temperature gradients [22, 23]) can also generate a $K$-dependence of the HBT radii [16]. Furthermore, the pion momentum spectra in particular are affected by resonance decay contributions, but only at small $K_\perp$. This may also affect the HBT radii in a $K_\perp$-dependent way [26]. The isolation of collective flow, in particular transverse flow, from the $K_\perp$-dependence of the HBT radii thus requires a careful study of these different effects.

We will here study this $K$-dependence of the HBT radius parameters within a simple analytical model for a finite thermalized source which expands both longitudinally and transversally. We will use the Yano-Koonin-Podgoretskii parametrization [27, 28, 17, 13] of the correlation function which, for sources with dominant longitudinal expansion, was shown to provide an optimal separation of the spatial and temporal aspects of the source [17, 13]. We point out that, at least in the absence of resonance decay contributions, it is a generic feature of all such thermal models that without transverse collective flow the YKP size parameters show perfect $M_\perp$-scaling, i.e. do not depend explicitly on the rest mass of the particles. Only the transverse gradients induced by a nonvanishing transverse flow can break this $M_\perp$-scaling. Thus accurate measurements of the YKP correlation radii with both pions and kaons can identify the presence of transverse collective flow in a rather model-independent way.

At the present moment it is still not absolutely clear to what extent this statement must be qualified when resonance decays are included. We have checked, however, that the $K_\perp$-independence of the the transverse HBT radius $R_\perp$ in the absence of transverse flow and other transverse gradients is not modified by resonance decays [29]. Thus, if there is no transverse flow, resonance decays cannot fake a transverse flow signal $R_\perp$.

## 2. The source model

For our study we use the model of Ref. [17] for an expanding thermalized source:

$$S(x,K) = \frac{M_\perp \cosh(\eta-Y)}{(2\pi)^3 \sqrt{2\pi(\Delta\tau)^2}} \exp\left[-\frac{K \cdot u(x)}{T(x)} - \frac{(\tau-\tau_0)^2}{2(\Delta\tau)^2} - \frac{r^2}{2R^2} - \frac{(\eta-\eta_0)^2}{2(\Delta\eta)^2}\right]. \quad (1)$$

Here $r^2 = x^2+y^2$, the spacetime rapidity $\eta = \frac{1}{2}\ln[(t+z)/(t-z)]$ and the longitudinal proper time $\tau = \sqrt{t^2-z^2}$ parametrize the spacetime coordinates $x^\mu$, with measure $d^4x = \tau\, d\tau\, d\eta\, r\, dr\, d\phi$. $Y = \frac{1}{2}\ln[(E_K+K_L)/(E_K-K_L)]$ and $M_\perp = \sqrt{m^2+K_\perp^2}$ parametrize the longitudinal and transverse components of the pair momentum $\vec{K}$. $T(x)$ is the freeze-out temperature, $R$ is the transverse geometric (Gaussian) radius of the source, $\tau_0$ its average freeze-out proper time, $\Delta\tau$ the mean proper time duration of particle emission, and $\Delta\eta$ parametrizes [16] the finite longitudinal extension of the source. The expansion flow velocity $u^\mu(x)$ is parametrized as

$$u^\mu(x) = (\cosh\eta \cosh\eta_t(r), \sinh\eta_t(r)\,\vec{e}_r, \sinh\eta \cosh\eta_t(r))\,, \quad (2)$$

with a boost-invariant longitudinal flow rapidity $\eta_l = \eta$ ($v_l = z/t$) and a linear transverse flow rapidity profile

$$\eta_t(r) = \eta_f \left(\frac{r}{R}\right) . \qquad (3)$$

$\eta_f$ scales the strength of the transverse flow. The exponent of the Boltzmann factor in (1) can then be written as

$$K \cdot u(x) = M_\perp \cosh(Y - \eta) \cosh \eta_t(r) - \vec{K}_\perp \cdot \vec{e}_r \sinh \eta_t(r) . \qquad (4)$$

For vanishing transverse flow ($\eta_f = 0$) the source depends only on $M_\perp$.

## 3. HBT radius parameters

From the source function (1) the correlation function is calculated via the relation

$$C(\vec{K}, \vec{q}) \approx 1 + \frac{\left|\int d^4x\, S(x,K)\, e^{iq \cdot x}\right|^2}{\left|\int d^4x\, S(x,K)\right|^2} = 1 + \frac{\left|\int d^4x\, S(x,K)\, e^{i\vec{q} \cdot (\vec{x} - \vec{\beta} t)}\right|^2}{\left|\int d^4x\, S(x,K)\right|^2} , \qquad (5)$$

(see [19, 30-32]). Here $q = p_1 - p_2$, $K = (p_1 + p_2)/2$ are the relative and average 4-momenta of the boson pair. The quality of the approximation in (5) is discussed in [16]. Since $p_1, p_2$ are on-shell and thus $K \cdot q = 0$, $q^0 = \vec{\beta} \cdot \vec{q}$ (with $\vec{\beta} = \vec{K}/K^0 \approx \vec{K}/E_K$) is not an independent variable and can be eliminated (second equality in (5)). Therefore the Fourier transform in (5) cannot be inverted without a model for $S(x, K)$. This is the reason for the model dependence of the interpretation of HBT correlation data mentioned in the Introduction.

We use a cartesian coordinate system with the $z$-axis along the beam direction and the $x$-axis along $\vec{K}_\perp$. Then $\vec{\beta} = (\beta_\perp, 0, \beta_l)$. We assume an azimuthally symmetric event sample and write the $C(\vec{q}, \vec{K})$ in the YKP parametrization [17, 13]:

$$C(\vec{K}, \vec{q}) = 1 + \exp\left[-R_\perp^2 q_\perp^2 - R_\parallel^2 \left(q_l^2 - (q^0)^2\right) - \left(R_0^2 + R_\parallel^2\right) (q \cdot U)^2\right]. \qquad (6)$$

Here $q_\perp^2 = q_x^2 + q_y^2$, and $R_\perp$, $R_\parallel$, $R_0$, $U$ are four $K$-dependent parameter functions. $U(\vec{K})$ is a 4-velocity with only a longitudinal spatial component:

$$U(\vec{K}) = \gamma(\vec{K}) \left(1, 0, 0, v(\vec{K})\right), \quad \text{with} \quad \gamma = \frac{1}{\sqrt{1 - v^2}} . \qquad (7)$$

Its value depends, of course, on the measurement frame. The "Yano-Koonin velocity" $v(\vec{K})$ can be calculated in an arbitrary reference frame from the second order space-time variances of the source $S(x, K)$ (for explicit expressions see Ref. [13]). It is, to a good approximation, the longitudinal velocity of the fluid element from which most of the particles with momentum $\vec{K}$ are emitted [17, 13]. For sources with boost-invariant longitudinal expansion velocity the YK-rapidity associated with $v(\vec{K})$ is linearly related to the pair rapidity $Y$ [13].

The other three YKP parameters do not depend on the longitudinal velocity of the observer system. Their physical interpretation is easiest in terms of coordinates measured in the frame where $v(\vec{K})$ vanishes. There they are given by [17]

$$R_\perp^2(\vec{K}) = \langle \tilde{y}^2 \rangle, \tag{8}$$

$$R_\parallel^2(\vec{K}) = \left\langle (\tilde{z} - (\beta_l/\beta_\perp)\tilde{x})^2 \right\rangle - (\beta_l/\beta_\perp)^2 \langle \tilde{y}^2 \rangle \approx \langle \tilde{z}^2 \rangle, \tag{9}$$

$$R_0^2(\vec{K}) = \left\langle (\tilde{t} - \tilde{x}/\beta_\perp)^2 \right\rangle - \langle \tilde{y}^2 \rangle/\beta_\perp^2 \approx \langle \tilde{t}^2 \rangle. \tag{10}$$

Here $\langle f(x) \rangle \equiv \int d^4x\, f(x)\, S(x,K) / \int d^4x\, S(x,K)$ denotes the ($K$-dependent) average over the source function $S(x,K)$, and $\tilde{x} \equiv x - \bar{x}(\vec{K})$ etc., where $\bar{x}(\vec{K})$ is the source point with the highest intensity at momentum $\vec{K}$. $R_\perp$, $R_\parallel$ and $R_0$ thus measure, approximately, the ($K$-dependent) transverse, longitudinal and temporal regions of homogeneity of the source in the local comoving frame of the emitter. The approximation in (9,10) consists of dropping terms which vanish in the absence of transverse flow and were found in [17] to be generically small even for finite transverse flow (see below). Note that it leads to a complete separation of the spatial and temporal aspects of the source.

Since in the absence of transverse flow the $\beta$-dependent terms in (9) and (10) vanish and the source itself depends only on $M_\perp$, it is clear that then all three YKP radius parameters show perfect $M_\perp$-scaling. Plotted as a function of $M_\perp$, they coincide for pion and kaon pairs. For nonvanishing transverse flow this $M_\perp$-scaling is broken by two effects: (1) The exponent of the thermal factor in (1) receives an additional contribution proportional to $K_\perp = \sqrt{M_\perp^2 - m^2}$. (2) The terms which were neglected in the second equalities of (9,10) are non-zero, and they also depend on $\beta_\perp = K_\perp/E_K$. Both effects induce an explicit dependence on the particle rest mass and destroy the $M_\perp$-scaling of the YKP size parameters.

## 4. Results

In Fig. 1 we show numerical results for the YKP correlation radii for the source (1), with the parameters $R = 3$ fm, $\tau_0 = 3$ fm/c, $\Delta\tau = 1$ fm/c, $\Delta\eta = 1.2$, and a constant temperature (no temperature gradients!) $T = 140$ MeV. The dashed lines refer to pion, the solid lines to kaon correlations, both at pair rapidity $Y = 0$ in the source c.m. frame. Resonance decays are not included (but see discussion above).

In the left column we show the HBT radii for a source without transverse expansion. Due to the absence of transverse temperature gradients, the transverse HBT radius $R_\perp$ is independent of $M_\perp$ and equal to the geometric Gaussian radius $R$ of the source. The strong $M_\perp$-dependence of the longitudinal radius $R_\parallel$ and the effective lifetime $R_0$ is caused by the strong longitudinal expansion with $v_l = z/t$.

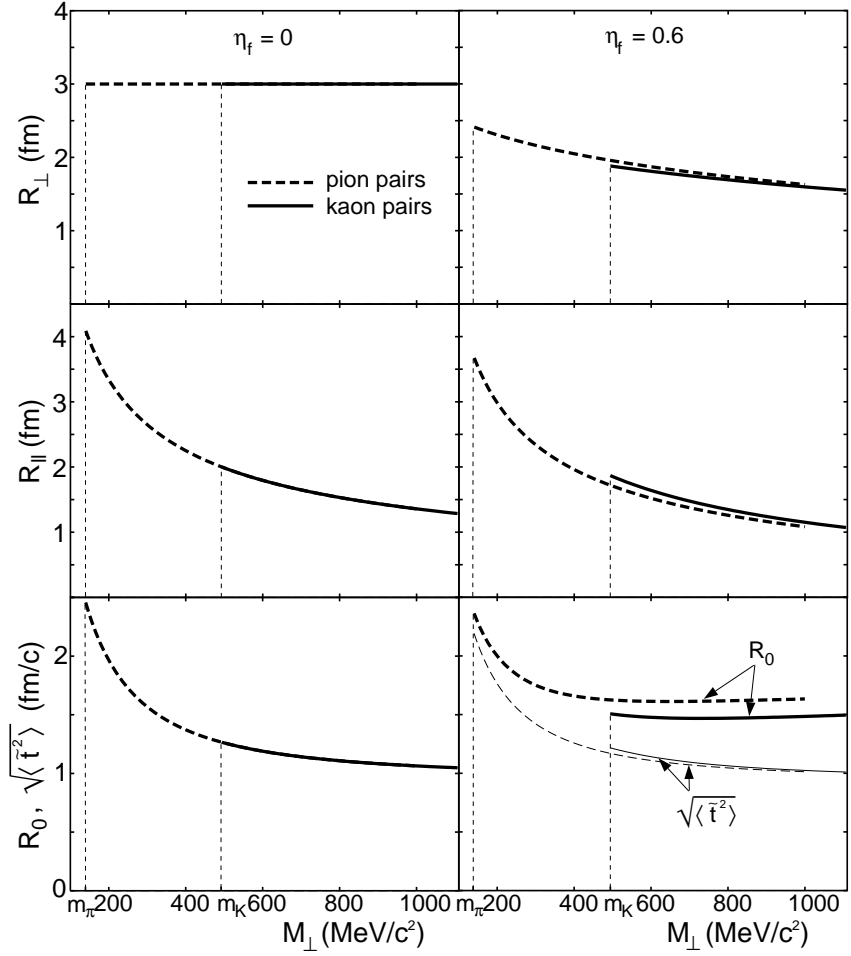

**Fig.1.** The YKP radii $R_\perp$, $R_\parallel$, and $R_0$, for pairs with $Y = 0$ as functions of $M_\perp = (m^2 + K_\perp^2)^{1/2}$. Dashed lines: pion correlations $(m = m_\pi)$; solid lines: kaon correlations $(m = m_K)$. Left column: no transverse expansion of source; right column: transverse expansion according to (3) with $\eta_f = 0.6$. For more discussion see text.

For large $M_\perp$, $R_0$ approaches the width $\Delta\tau$ of the $\tau$-distribution in (1), but for small $M_\perp$ it is significantly larger. Over the finite longitudinal region $R_\parallel$ from which the correlator receives contributions, the proper time hyperbola at $\tau=\tau_0$ covers a finite range $\Delta t$ (in any fixed frame, in particular in the source rest frame). $R_0$ measures this $\Delta t$ plus the Gaussian width $\Delta\tau$. Since $R_\parallel$ decreases for large $M_\perp$, so does $\Delta t$, and only $\Delta\tau$ survives in the limit.

For vanishing transverse flow $\eta_f = 0$, all three YKP radii show perfect $M_\perp$-scaling. If the source expands longitudinally like ours, $R_\parallel$ and $R_0$ are thus smaller for kaon pairs than for low-momentum pion pairs, even without resonance decays. Due to the smaller $R_\parallel$ for kaons, they cannot probe the longitudinal variation $\Delta t$ of time along the freeze-out surface, and $R_0$ for kaons essentially measures only the Gaussian widths $\Delta\tau$ of the proper emission time distribution.

The right column corresponds to a transversally expanding source with $\eta_f$=0.6. This value is probably at the upper edge of the range required to explain present heavy ion data [4, 7, 33]. One sees three types of effects: (a) Now also the transverse radius $R_\perp$ develops an $M_\perp$-dependence. The $M_\perp$ slope is directly related to the magnitude of $\eta_f$ [29, 34]. (b) The $M_\perp$-scaling of the YKP radii is broken. The effect is not very big, but it goes in *opposite* directions for $R_\parallel$ and the two other radius parameters. This will be helpful in disentangling flow from possible resonance decay effect which are stronger for pions than for kaons and always tend to increase all three radius parameters. (c) $R_0$ receives contributions from the last two terms in the middle expression in Eq. (10) and no longer just measures the effective lifetime $\sqrt{\langle\tilde{t}^2\rangle}$. (The corresponding contributions to $R_\parallel$ in Eq. (9) still vanish, due to $z$-symmetry at $Y = 0$.) The scale-breaking effects from the correction terms are in fact stronger and of opposite sign relative to the lifetime $\sqrt{\langle\tilde{t}^2\rangle}$. The relative difference between $R_0$ and $\sqrt{\langle\tilde{t}^2\rangle}$ is expected to decrease for larger systems with larger values of $\tau_0$ and $\Delta\tau$. In this sense $R_0$ is still a good measure for the effective lifetime, actually more so for kaons than for pions.

## Acknowledgement

This work was supported by grants from DAAD, DFG, NSFC, BMBF and GSI. We gratefully acknowledge discussions with H. Appelshäuser, S. Chapman, D. Ferenc, M. Gaździcki, and P. Seyboth. U.H. expresses his thanks to B. Müller and the Physics Department at Duke University for their warm hospitality.

## Notes

[a] On sabbatical leave from Institut für Theoretische Physik, Univ. Regensburg.
[b] On leave of absence from Institute of Particle Physics, Hua-Zhong Normal University, Wuhan, China.